\documentclass[twocolumn,showpacs,preprintnumbers,amsmath,amssymb,superscriptaddress,floatfix]{revtex4}
\usepackage{graphicx}
\usepackage{amsmath}
\usepackage{bm}
\usepackage{color}

\usepackage{graphicx}
\usepackage{dcolumn}
\usepackage{bm}

\begin{document}

\title{Sonic Mach Cones Induced by Fast Partons in a Perturbative Quark-Gluon Plasma}

\author{R. B. Neufeld}
   \affiliation{Department of Physics, Duke University, Durham, NC 27708, USA}
\author{B. M\"uller}
   \affiliation{Department of Physics, Duke University, Durham, NC 27708, USA}
\author{J. Ruppert}
   \affiliation{Institut f\"ur Theoretische Physik, J. W. Goethe-Universit\"at, D-60054 Frankfurt am Main, Germany}
 \affiliation{
{Department of Physics, McGill University, Montreal, Quebec, H3A 2T8, Canada}}

\date{\today}

\begin{abstract}

We derive the space-time distribution of energy and momentum deposited by a fast parton traversing a weakly coupled quark-gluon plasma by treating the fast parton as the source of an external color field perturbing the medium.  We then use our result as a source term for the linearized hydrodynamical equations of the medium.  We show that the solution contains a sonic Mach cone and a dissipative wake if the parton moves at a supersonic speed.
\end{abstract}

\pacs{12.38.Mh,25.75Ld,25.75.Bh}

\maketitle

An interesting problem in the physics of deconfined strongly interacting matter -- the quark-gluon plasma -- is to calculate the response of the medium to the passage of a fast parton, i.~e., a parton with velocity approaching the speed of light.  Fast partons are created experimentally in high-energy collisions of two nuclei when two energetic partons of the nuclear wave functions scatter at a large angle and acquire a large transverse momentum relative to the beam direction.  If the interaction happens near the nuclear surface, one parton (commonly called the ``trigger jet'') rapidly leaves the medium and decays into a jet of hadrons which are cleanly observed in the detectors, while the second parton (commonly called the ``back-jet'') propagates with high velocity through the medium and deposits energy and momentum along the way in a process known as jet quenching (see, e.~g., \cite{Baier:2000mf,Jacobs:2004qv}).  Experimental measurements \cite{Adams:2005ph,Adler:2005ee} of di-hadron correlation functions show a double peak structure in the back-jet distribution which is consistent with the formation of a Mach cone-shaped emission pattern.

The question of how the energy and momentum deposited by the back-jet affects the bulk behavior of an evolving quark-gluon plasma has been examined by several authors (see, e.~g., \cite{CasalderreySolana:2004qm,Stoecker:2004qu,Satarov:2005mv,Ruppert:2005uz,Renk:2005si,Chaudhuri:2005vc,Friess:2006fk,Casalderrey-Solana:2007km,Chesler:2007sv}).  Using a schematic source in a two-dimensional hydrodynamics simulation of an expanding quark-gluon plasma, Chaudhuri and Heinz failed to find the formation of a Mach cone except for extreme values of the energy deposition \cite{Chaudhuri:2005vc}.  On the other hand, Casalderrey-Solana {\it et al.} \cite{CasalderreySolana:2004qm} showed that if one couples an appropriately chosen supersonic sound source to a linearized hydrodynamical evolution one obtains a propagating Mach cone.  \textcolor{red}{In the above cases the explicit form of the source term expected from a supersonic parton moving through a QCD plasma was not addressed.}

Here we will derive \textcolor{red}{the magnitude and shape of this source term, valid at distances below the Debye screening length}, by calculating the effect of the color field of a fast parton on a perturbative quark-gluon plasma.  We then solve the linearized hydrodynamical equations for a static quark-gluon plasma and analyze how the perturbation propagates through the medium.  For sound propagation we will use values of the sound attenuation length, $\Gamma_s$,  which lie in the range compatible with perturbation theory.  These include the value obtained from the leading-order perturbative result for the shear viscosity $\eta$ \cite{Arnold:2003zc} and the value for $\eta$ obtained by including three-body scattering processes in the transport cross section \cite{Xu:2007ns,Xu:2007jv}.  We note that the latter value is more consistent with shear viscosity allowed by the RHIC data \cite{Romatschke:2007mq}.

As explained above, we consider the fast parton as the source of an external color field that interacts with the medium.  Asakawa {\it et al.} \cite{Asakawa:2006jn} showed that in the presence of soft color fields the color singlet parton distribution in a perturbative quark-gluon plasma obeys the following dissipative Vlasov-Boltzmann equation:
\begin{equation}\label{finalVB}
\left[\frac{\partial}{\partial t} + \frac{{\mathbf p}}{E}\cdot {\bm\nabla}_x  - \nabla_{p_i} {D_{ij}}({\mathbf p},t)\nabla_{p_j}\right]f({\mathbf x},{\mathbf p},t) = C[f]
\end{equation}
where $f({\mathbf x},{\mathbf p},t)$ is the ensemble averaged phase space distribution of medium partons, ${\mathbf p}/E$ is the velocity of a parton with momentum ${\mathbf p}$ and energy $E$, and
\begin{equation}\label{dupree}
{D_{ij}}({\mathbf p},t) = \int_{-\infty}^{t}dt'F_i({\mathbf x},t)F_j({\mathbf x}',t') ,
\end{equation}
where $F_i({\mathbf x},t)$ is the color Lorentz force on a medium particle:
\begin{equation}\label{lorentz}
F_i({\mathbf x},t) = g {Q^{a}(t)}\left({E^{a}_i}({\mathbf x},t) + ({{\mathbf v}\times{\mathbf B}})^{a}_i({\mathbf x},t)\right).
\end{equation}
Here we consider the color fields in (\ref{lorentz}) to be generated by the fast parton.  For a parton moving with velocity ${\mathbf u}$ with respect to the medium these fields, to lowest order in the coupling constant $g$, have the Fourier representation
\begin{equation}\label{efield}
{\mathbf E}^a({\mathbf x},t) = \frac{i g {Q_p^a}}{(2 \pi)^3}\int d^4k \text{ }e^{-ik\cdot x}\frac{(\omega {\mathbf u} - {\mathbf k})\delta(\omega - {\mathbf k} \cdot {\mathbf u})}{k^2 - \omega^2}
\end{equation}
\begin{equation}\label{bfield}
{\mathbf B}^a({\mathbf x},t) = \frac{i g {Q_p^a}}{(2 \pi)^3}\int d^4k \text{ }e^{-ik\cdot x}\frac{({\mathbf k}\times{\mathbf u})\delta(\omega - {\mathbf k} \cdot {\mathbf u}))}{k^2 - \omega^2},
\end{equation}
where $k^\mu = (\omega,{\mathbf k})$.  In (\ref{efield}, \ref{bfield}) we have neglected the screening of the color field by the medium.
We will incorporate the effect of color screening and short-distance quantum effects by appropriate infrared and ultraviolet cut-offs.

The hydrodynamical equations for the medium are obtained in the usual Chapman-Enskog approach by taking moments of the evolution equation (\ref{finalVB}).  We assume that the medium is in local thermal equilibrium, with an energy-momentum tensor given by
\begin{equation}\label{basictensor}
T^{\mu \nu} = (\epsilon + p)u^\mu u^\nu - p g^{\mu \nu} - \eta(\nabla^\mu u^\nu + \nabla^\nu u^\mu - \frac{2}{3} \Delta^{\mu\nu} \partial_\alpha u^\alpha)
\end{equation}
where $u^\mu = \gamma(1,{\mathbf v})$ denotes the medium four-velocity, $\Delta^{\mu\nu} = u^\mu u^\nu - g^{\mu\nu}$, $\nabla^{\mu} = \Delta^{\mu\nu}\partial_\nu$, and $\eta$ is the shear viscosity.  We evaluate each term by boosting to a frame co-moving with our volume element and then exploiting the assumption of local thermal equilibrium.  Introducing the notation
\begin{equation}\label{sourceterm}
J^{\nu} = \int \frac{d {\mathbf p} \text{ }p^\nu}{(2 \pi)^3} (\nabla_{p_i} {D_{ij}}({\mathbf p},t)\nabla_{p_j}f({\mathbf x},{\mathbf p},t)) ,
\end{equation}
we find that the resulting equations of motion for the medium evolution are
\begin{equation}\label{sourcehydro}
\partial_\mu T^{\mu \nu} = J^{\nu} ,
\end{equation}
where $J^\nu$ represents a source term due to the interaction of the medium with the passing fast parton.

In order to solve the hydrodynamical equations (\ref{sourcehydro}), we assume that the energy and momentum density deposited by the parton is small compared to the equilibrium energy density of the medium.  This assumption permits us to linearize the hydrodynamical equations following the approach introduced in \cite{CasalderreySolana:2004qm}.  We write the perturbations of the energy-momentum tensor resulting from the source in an otherwise static medium as $\delta\epsilon = \delta T^{00}$ and  ${\mathbf g}$ with $g_i = \delta T^{0i}$.  The solutions of the linearized hydrodynamical equations can then be expressed in momentum space as
\begin{eqnarray}
\delta\epsilon ({\mathbf k},\omega) &=& \frac{i k J_L({\mathbf k},\omega)  + J^0({\mathbf k},\omega)(i \omega -  \Gamma_s k^2)}{\omega^2 -  c_s^2 k^2 + i \Gamma_s \omega k^2},
\label{one} \\
{\mathbf g}_L ({\mathbf k},\omega) &=& \frac{ i \omega \hat{k} J_L({\mathbf k},\omega)+ i c_s^2 {\mathbf k} J^0({\mathbf k},\omega)}{\omega^2 -  c_s^2 k^2 + i \Gamma_s \omega k^2},
\label{two} \\
{\mathbf g}_T ({\mathbf k},\omega) &=& \frac{i{\mathbf J}_T({\mathbf k},\omega)}{\omega +  \frac{3}{4}i \Gamma_s k^2}
\label{three}
\end{eqnarray}
where $\epsilon_0$ and $p_0$ are the unperturbed energy density and pressure, respectively, $c_s$ denotes the speed of sound, and $\Gamma_s \equiv \frac{4 \eta }{3(\epsilon_0 + p_0)} =  \frac{4 \eta }{3s_0 T}$ is the sound attenuation length.  In the above equations we have divided the source and perturbed momentum density vectors into transverse and longitudinal parts: ${\mathbf g} = {\mathbf g}_L + {\mathbf g}_T$ and ${\mathbf J} = \hat{\mathbf k} J_L + {\mathbf J}_T$, with $\hat{\mathbf k}$ denoting the unit vector in the direction of ${\mathbf k}$.  We note that (\ref{three}) is the diffusion equation and does not describe sound propagation.  If a Mach cone appears it will be found in the dynamics of eqs.~(\ref{one}, \ref{two}).

In order to proceed further we need to evaluate the source term (\ref{sourceterm}).  We consider a thermal plasma of massless gluons with the unperturbed distribution
\begin{equation}\label{boseeinstein}
f_0({\mathbf x},{\mathbf p},t) = \frac{2 ({N^{2}_c}-1)}{e^{\beta p^0}-1}
\end{equation}
where $p^0 = E = |{\mathbf p}|$ and $N_c=3$ is the number of colors.  Ignoring the time dependence of ${Q^{a}(t)}$ in (\ref{lorentz}) we eventually find that
\begin{equation}\label{reducednot}
\begin{split}
J^{0}({\mathbf x},t) = \frac{i m_{\rm D}^2}{(2 \pi)^8}\int d^4k d^4k' e^{i ({\mathbf k} + {\mathbf k'})\cdot {\mathbf x} - i(\omega + \omega')t}\times \\
\int d \hat{ \mathbf v} \frac{ \left(\hat{{\mathbf v}}\cdot{\mathbf E}^a(k')\right) \left(\hat{{\mathbf v}}\cdot{\mathbf E}^a(k)\right) }{4 \pi(\omega' - {\mathbf k'}\cdot \hat{ \mathbf v} + i \epsilon)}
\end{split}
\end{equation}
and
\begin{equation}\label{reducedk}
\begin{split}
J^{k}({\mathbf x},t) = \frac{i m_{\rm D}^2}{(2 \pi)^8}\int d^4k d^4k' e^{i ({\mathbf k} + {\mathbf k'})\cdot {\mathbf x} - i(\omega + \omega')t}\times \\
\int d \hat{ \mathbf v} \frac{ \left(\hat{{\mathbf v}}\cdot{\mathbf E}^a(k')\right) \left({E^{a}_k}(k) + ({\hat{ \mathbf v}\times{\mathbf B}})^{a}_k(k)\right)}{4 \pi(\omega' - {\mathbf k'}\cdot \hat{ \mathbf v} + i \epsilon)}
\end{split}
\end{equation}
where the fields ${\mathbf E}^a(k)$ and ${\mathbf B}^a(k)$ are given by (\ref{efield}) and (\ref{bfield}), respectively.  The Debye mass for gluons is given by $m_{\rm D} = gT$.  If one includes massless quarks in the medium the expressions (\ref{reducednot} and  \ref{reducedk}) remain unchanged except that the quark contribution to the Debye mass needs to be included in ${m_{\rm D}}$.  The longitudinal part of ${\mathbf J}$ is obtained by multiplication with $\hat{\mathbf k}$: $ J_L =\hat{{\mathbf k}}\cdot {\mathbf J}$.  The source distributions can be analytically evaluated in the ultra-relativistic limit $\gamma = (1-u^2)^{-1/2} \gg 1$. After a lengthy calculation one obtains for a parton moving in the positive $z$-direction with velocity $u$:
\begin{eqnarray}
J^0(\rho,z,t) &=& d(\rho,z,t) \gamma u^2 \left( 1 - \frac{\gamma u(z-ut)}{\rho} \right) ,
\label{J0}
\\
J_z(\rho,z,t) &=& u J^0(\rho,z,t) - d(\rho,z,t) u^2 \frac{z-ut}{\rho} ,
\label{Jz}
\\
{\mathbf J}_\perp(\rho,z,t) &=& - d(\rho,z,t) u^2 \frac{\mathbf x_\perp}{\rho} ,
\label{Jperp}
\end{eqnarray}
where we introduced the notation ${\mathbf x_\perp} = (x,y)$ and $\rho = |{\mathbf x_\perp}|$.  The function $d$ is given by
\begin{equation}
d(\rho,z,t) = \frac{\alpha_s Q_p^2 m_{\rm D}^2}{8\pi[\rho^2 + \gamma^2(z-ut)^2]^{3/2}} ,
\label{d(x)}
\end{equation}
encoding the Lorentz contracted field configuration of the fast parton.  A detailed derivation of the source term, including color screening by the medium, is presented in \cite{Neufeld:2008hs}.  Integrating over all space and introducing infrared and ultraviolet cut-offs for the $\rho$-integration, we obtain the differential energy loss
\begin{equation}\label{dE_dx}
-\frac{dE}{dx} = \int d^3x J^0(x) = \frac{\alpha_s}{2} Q_p^2 m_{\rm D}^2  \ln \frac{\rho_{\rm max}}{\rho_{\rm min}} .
\end{equation}
For $\rho_{\rm max} = 1/m_{\rm D}$ and $\rho_{\rm min} = 1/(2\sqrt{E_p T})$, where $E_p$ is the energy of the fast parton, one recovers the standard leading-logarithmic result \cite{Thoma:1991ea} for elastic energy loss in the medium.  A stronger than logarithmic dependence of the energy loss on $E_p$, found in some other approaches, may be effectively described by a parametric dependence of $\alpha_s Q_p^2$ on $E_p$.

Since hydrodynamics is only valid at distances large compared to the mean free path, and in a weakly coupled plasma the mean free path is parametrically large compared to the color screening length, the source term generated by an energetic parton is, in first  approximation, point-like.  In this spirit, the source term (\ref{J0}-\ref{d(x)}) derived here can be thought of as a sophisticated representation of a delta function.  However, it also contains, in the sense of effective field theory, an infinite series of higher derivative contributions, which produce sub-leading corrections.

At leading order in the gradient expansion, both the amplitude and structure of the Mach cone produced by the projectile are uniquely determined by the rate of elastic energy loss together with the value of viscosity. The source term derived and used here provides a model of the sub-leading momentum dependence of the source term for the special case of a weakly coupled plasma, which approximately captures the physics at the Debye screening scale, but ignores physics at longer sub-hydrodynamic length scales \cite{Ref}.

To calculate the propagating disturbance of the medium by the projectile we insert (\ref{reducednot}) and (\ref{reducedk}) into eqs.~(\ref{one} -- \ref{three}).  For a relativistic parton, propagating with velocity ${\mathbf u} = (0,0,u)$ and position ${\mathbf x} = (0,0,ut)$ we obtain the following expressions for the quantities (\ref{one} -- \ref{three}), Fourier transformed back to space-time:
\begin{widetext}
\begin{equation}
\label{rone}
\delta \epsilon({\mathbf x},t) = \frac{i u^2 g^2 Q_p^2 {m^{2}_D}}{32 \pi^2(2\pi)^3(u^2 - c_s^2)} \int d{\mathbf k} d{\mathbf x'} e^{i {\mathbf k}\cdot({\mathbf x}-{\mathbf x'}) - i u k_z t}\left(\frac{-{\mathbf k}\cdot {\mathbf x'} + (2 u k_z + i \Gamma_s k^2)\left(\rho' \gamma - z' \gamma^2 u\right)}{(k_z^2 - \lambda^2 k_T^2 + i\sigma)\rho' (\rho'^2 + \gamma^2 z'^2)^{3/2}}\right),
\end{equation}
\begin{equation}\label{rtwo}
{\mathbf g}_L({\mathbf x},t) = \frac{i u^2 g^2 Q_p^2 {m^{2}_D}}{32 \pi^2(2\pi)^3(u^2 - c_s^2)} \int d{\mathbf k} d{\mathbf x'} e^{i {\mathbf k}\cdot({\mathbf x}-{\mathbf x'}) - i u k_z t}{\mathbf k} \left(\frac{-u k_z{\mathbf k}\cdot {\mathbf x'} + ((u k_z)^2 + c_s^2 k^2)\left(\rho' \gamma - z'\gamma^2 u\right)}{k^2(k_z^2 - \lambda^2 k_T^2 + i\sigma)\rho' (\rho'^2 + \gamma^2 z'^2)^{3/2}}\right)
\end{equation}
\begin{equation}\label{rthree}
{\mathbf g}_T({\mathbf x},t) = \frac{1}{(2\pi)^3} \int d{\mathbf k} d{\mathbf x'} e^{i {\mathbf k}\cdot({\mathbf x}-{\mathbf x'}) - i u k_z t} \frac{{\mathbf J}({\mathbf x'})k^2 - {\mathbf k} ({\mathbf k}\cdot{\mathbf J}({\mathbf x'}))}{k^2(-i u k_z + \frac{3}{4}\Gamma_s k^2)}
\end{equation}
\end{widetext}
where $\lambda^2 = c_s^2/(u^2-c_s^2)$, $\sigma = \Gamma_s u (\lambda^2/c_s^2) k_z(k_T^2 + k_z^2)$, and $Q_p^2$ is the Casimir operator for the color charge of the projectile, which is $4/3$ for a quark and $3$ for a gluon.   In our evaluation of (\ref{rone}-\ref{rthree}) we perform five of the six integrals analytically leaving the final integral over $k_T \equiv (k_x^2 + k_y^2)^{1/2}$ to be done numerically.  In (\ref{rone}, \ref{rtwo}) we perform the integral over $k_z$ using contour integration and find poles at (among other places) $k_z = \pm (k_T^2 \lambda^2 \mp i |\sigma|)^{1/2}$, where $|\sigma|$ is itself a function of $k_z$.  We approximate $\sigma(k_z)$ by $\sigma(\pm k_T \lambda)$ when evaluating the residues at these poles, which is permitted at momentum scales for which the sound attenuation is small ($k_T \ll c_s^2/\Gamma_s$).   We note that our expression contains the term $(\rho'^2 + \gamma^2 z'^2)^{-3/2}$ which is (up to appropriate normalization) a nascent delta function in $z'$ in the limit $\gamma \rightarrow \infty$.  With this insight, and noting that we are working in the ultrarelativistic limit, we expand the $z'$ dependence of the exponent to first order to obtain analytically manageable expressions.  An ultraviolet cut-off of the order of $\Gamma_s$ in the $\rho'$ integration and an appropriate infrared cut-off in the $z'$ integration are used to regularize the logarithmic divergences when necessary.

We now present and discuss the numerical results obtained from our evaluation of (\ref{rone}-\ref{rthree}) for the case of a gluon moving along the positive $z$ axis with velocity $u = 0.99955 c$ ($\gamma\approx 33$).  As mentioned in the beginning, we will use two different values of the sound attenuation length, $\Gamma_s$, which have been calculated perturbatively and compare the results.  The first value is determined from the leading order result for $\eta/s$ obtained by Arnold {\em et al.} \cite{Arnold:2003zc}.  This result includes only binary ($2\rightarrow 2$ and $1\rightarrow 2$) processes and gives $\eta/s = 0.48$ for a gluonic plasma with $\alpha_s = 0.3$.  The second value for $\eta/s$ is the one obtained by Xu and Greiner by going beyond leading order in the diluteness of the medium \cite{Xu:2007aa}.  Their calculation, including $3\leftrightarrow 2$ scattering processes, found $\eta/s = 0.13$ for a gluonic plasma with $\alpha_s = 0.3$ \cite{Xu:2007ns}. Even smaller values of $\eta/s$ may be compatible with the assumption of a perturbative medium, if the viscosity is lowered by anomalous contributions \cite{Asakawa:2006tc}. Finally, we have chosen $T = 350$ MeV as the temperature of the medium and use $c_s = c/\sqrt{3}$ for the speed of sound.

\begin{figure*}
\centerline{
\includegraphics[width = 0.42\linewidth]{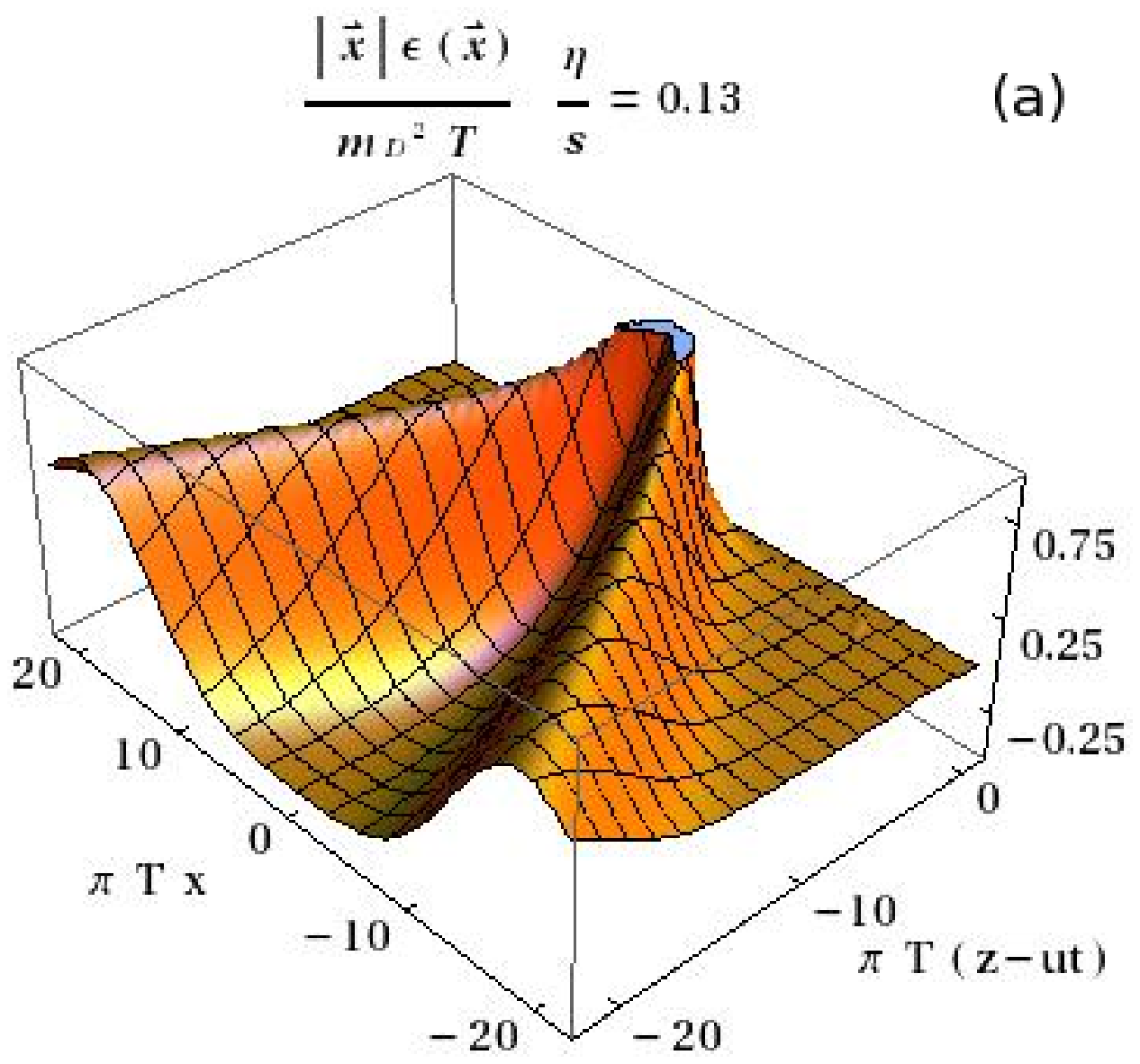} \hskip0.08\linewidth
\includegraphics[width = 0.42\linewidth]{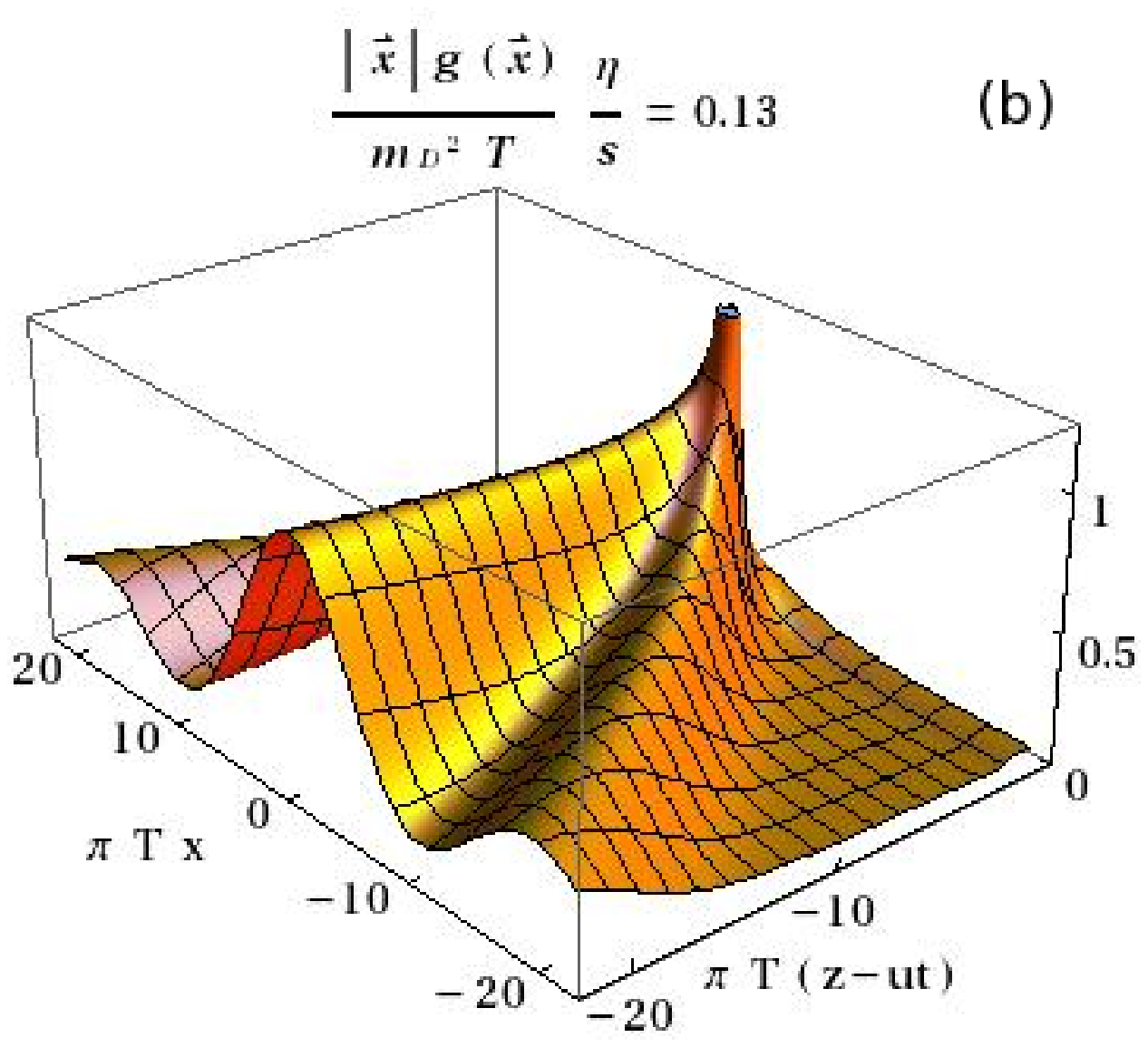}}
\caption{(Color online) Plots of the scaled perturbed energy density $|{\mathbf x}|\delta\epsilon({\mathbf x})/(m_{\rm D}^2 T)$ (left), and momentum density $|{\mathbf x}|{\mathbf g}({\mathbf x})/(m_{\rm D}^2 T)$ (right), excited by a gluon moving along the positive $z$ axis at position $ut$ and speed $u = 0.99955 c$ for $\eta/s = 0.13$.  The results, which are cylindrically symmetric around the $z$ axis, are shown on the plane ${\mathbf x} = (x,0,z-ut)$. The energy and momentum densities have been multiplied by the distance $|{\mathbf x}|$ from the source to compensate for the geometric dilution effect of the cone.  The values chosen for the parameters $m_{\rm D}$, $\alpha_s$, $T$, $c_s$, and $\Gamma_s$ are disussed in the text.}
\label{visc_13}
\end{figure*}

\begin{figure*}
\centerline{
\includegraphics[width = 0.42\linewidth]{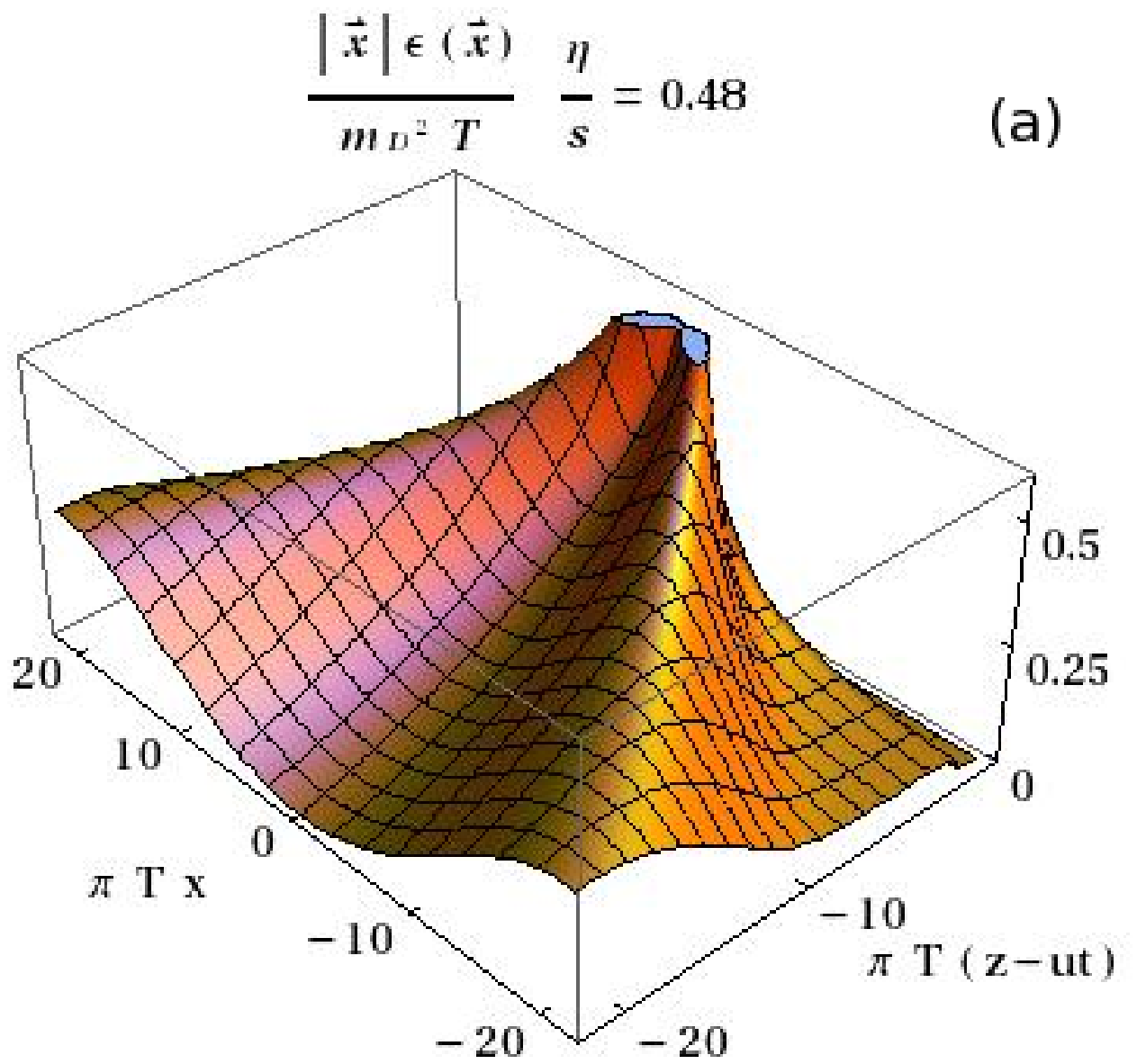} \hskip0.08\linewidth
\includegraphics[width = 0.42\linewidth]{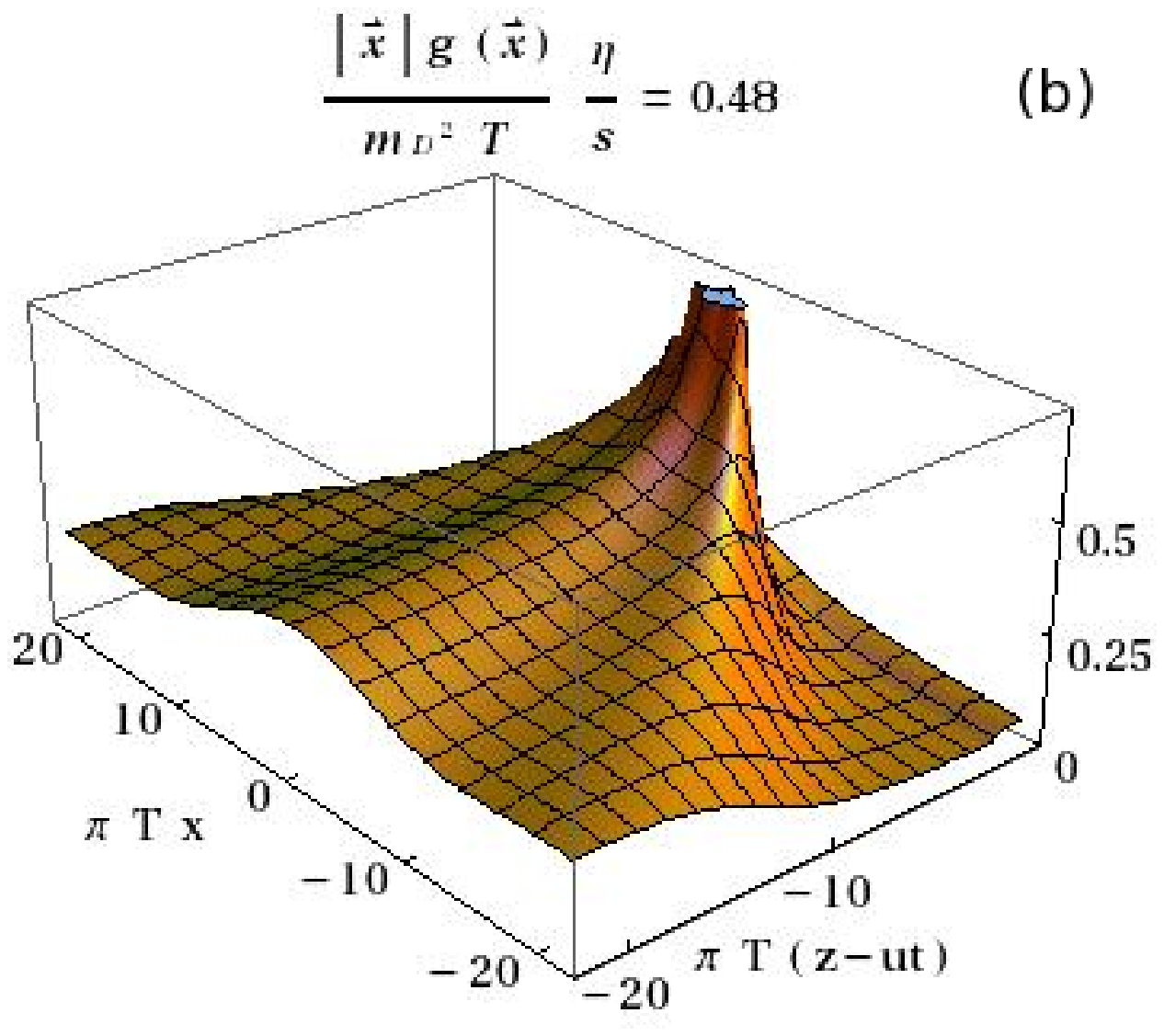}}
\caption{(Color online) The same as in Figure \ref{visc_13} but with $\eta/s = 0.48$.  Just as in Figure \ref{visc_13} the energy density wave excited in the medium develops the shape of a Mach cone.  Here, however, the distribution is more damped and spread out.}
\label{visc_48}
\end{figure*}

Figures \ref{visc_13} and \ref{visc_48} show the scaled energy density (left) and momentum density (right) deposited in the medium by the projectile along the plane $(x,0,z-ut)$ for $\eta/s = 0.13$ and $\eta/s = 0.48$, respectively.  In both cases the energy density wave excited in the medium is seen to develop the shape of a Mach cone whose intensity is peaked near the source.  At growing distance from the source the Mach cone slowly weakens and broadens.  As one would expect, the cone structure is more pronounced for the smaller value of $\eta/s$.  It is instructive to compare the wavelength below which sound propagation is strongly damped, $\lambda_c = 2\pi\Gamma_s/c_s$, with the extent of the sound source, $m_{\rm D}^{-1} \approx 0.3$ fm. For $\eta/s = 0.48$ we have $\lambda_c/4 \approx 1.0$ fm, while for $\eta/s = 0.13$ we find $\lambda_c/4 \approx 0.25$ fm, suggesting that the coupling of the source term to the sound mode becomes increasingly ineffective for $\eta/s \gg 0.15$.

\begin{figure*}
\centerline{
\includegraphics[width = 0.42\linewidth]{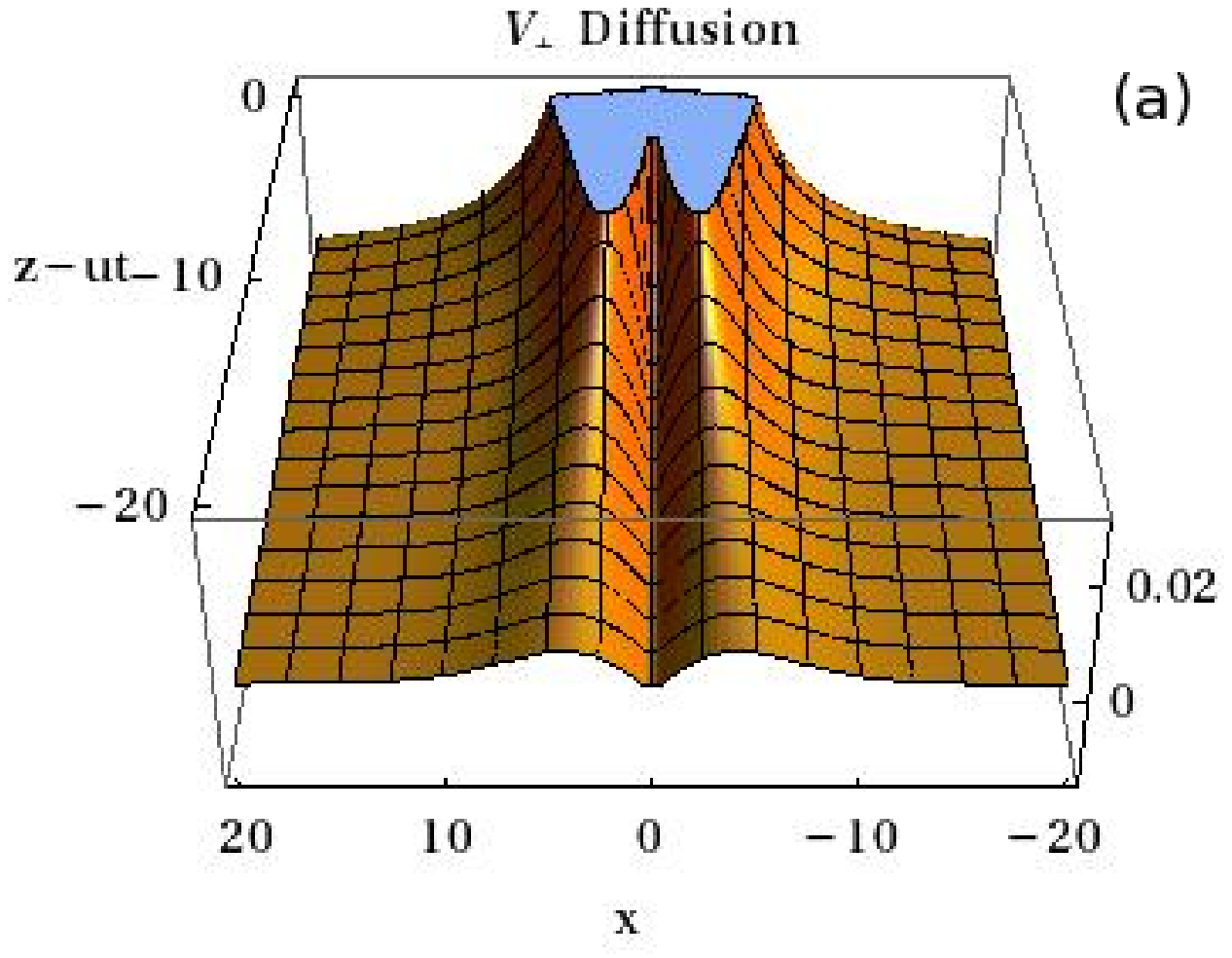} \hskip0.08\linewidth
\includegraphics[width = 0.42\linewidth]{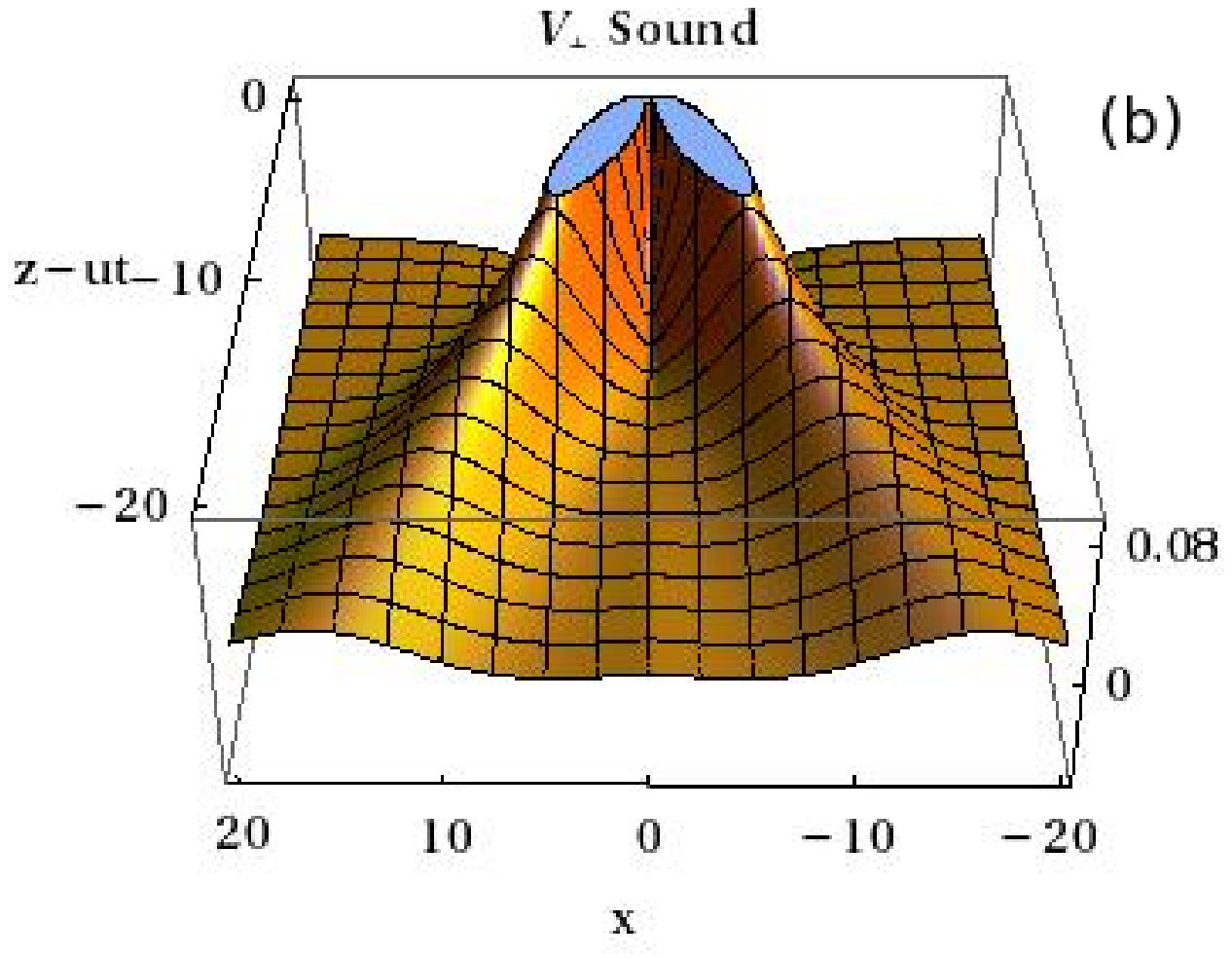}}
\centerline{
\includegraphics[width = 0.42\linewidth]{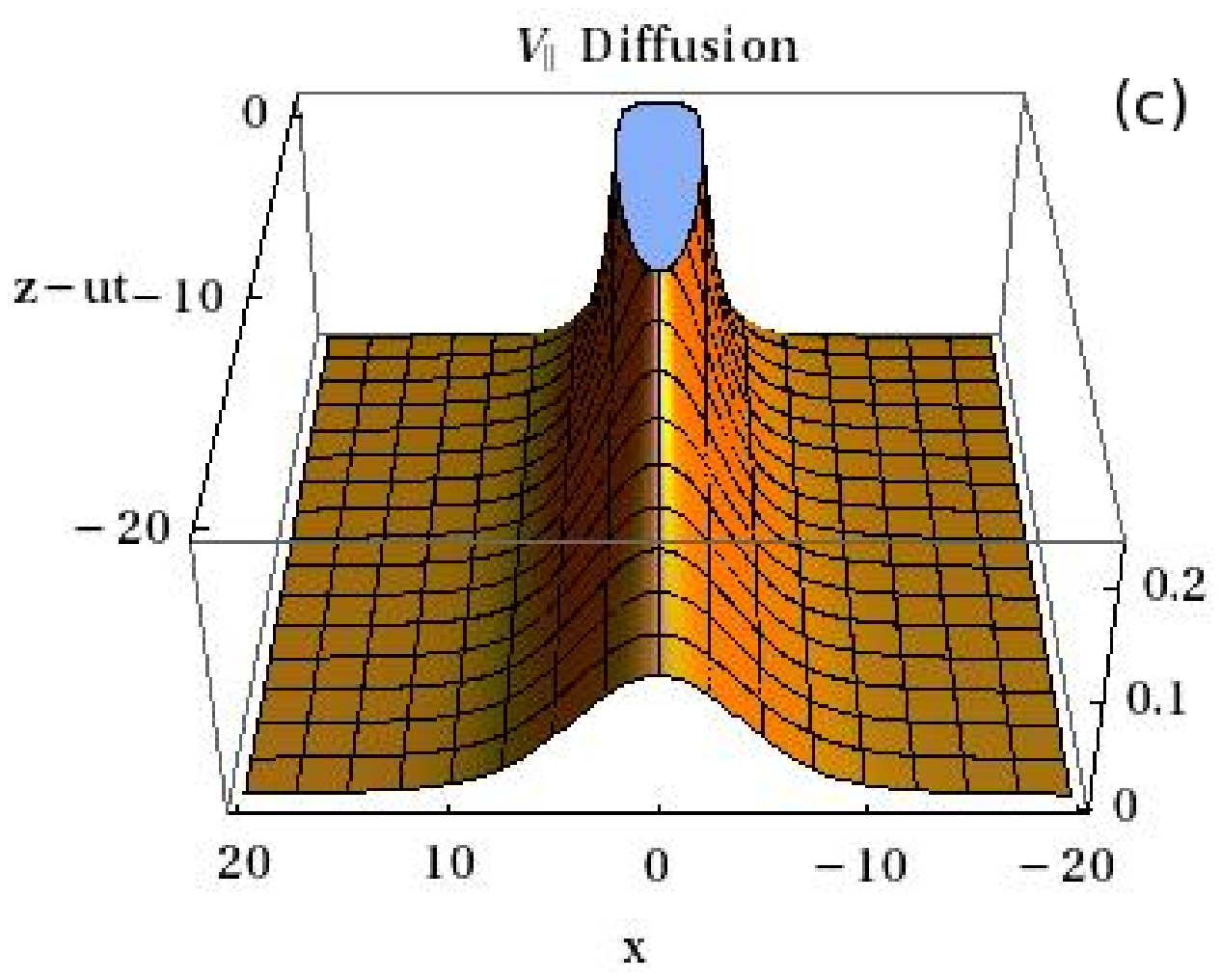} \hskip0.08\linewidth
\includegraphics[width = 0.42\linewidth]{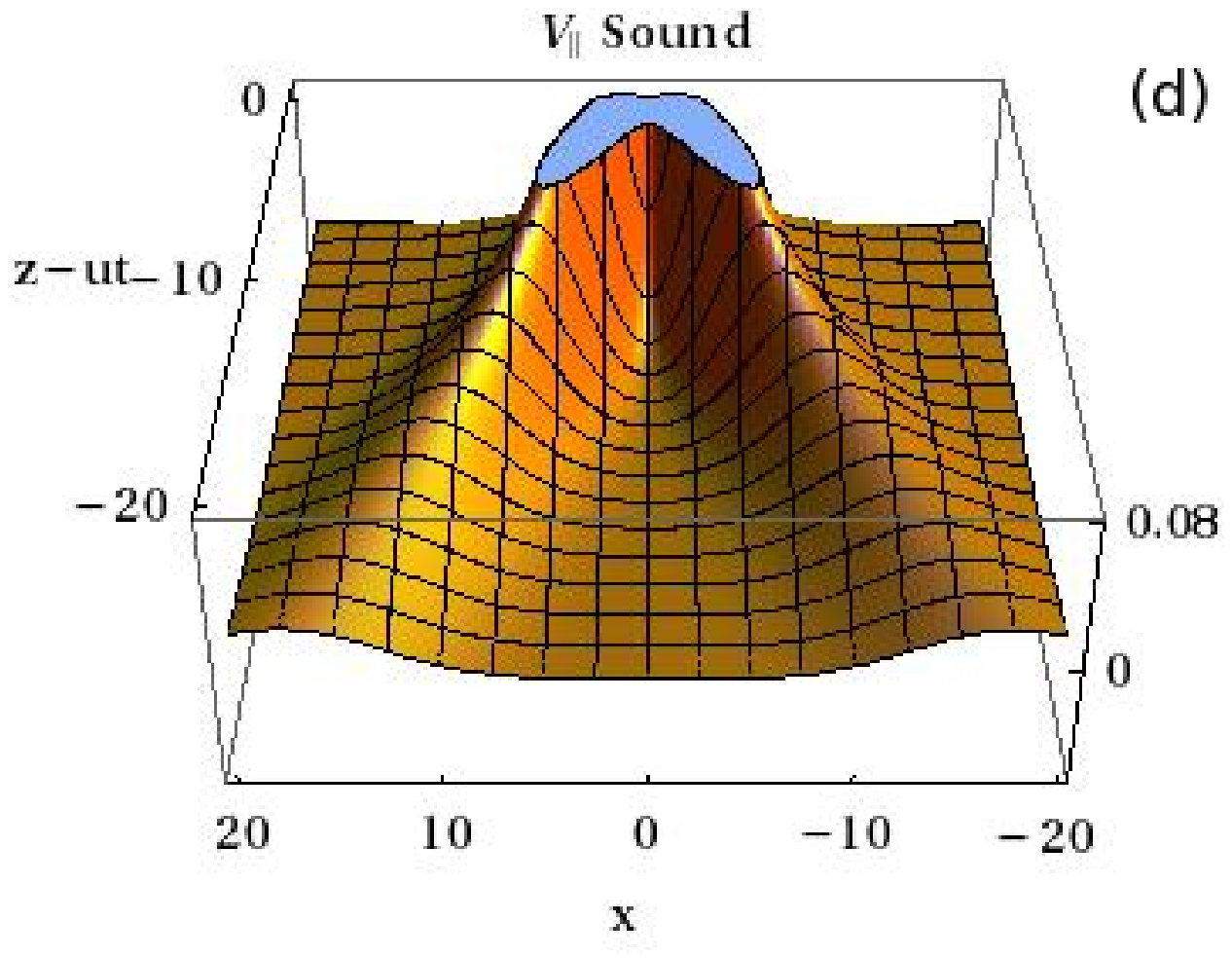}}
\caption{(Color online) Plots of the velocity field for the same scenario as in Figure \ref{visc_13}.  The two lower plots indicate the parallel (i.e. moving with the source) components of the induced velocity flows for both the sound equation (\ref{rtwo}) and the diffusion equation (\ref{rthree}) while the two upper plots are for the perpendicular components.  The axes are plotted in units of inverse GeV.}
\label{velocity}
\end{figure*}

We next consider the velocity fields induced by the moving source which are given by ${\mathbf v} = {\mathbf g}/\epsilon_0$ where $\epsilon_0 \approx 10 \text{ GeV}/\text{fm}^3$ is the energy density of a plasma of massless gluons at $T = 350$ MeV.  We plot the induced velocity flow in Fig.~\ref{velocity} for both the contribution from the sound equation (\ref{rtwo}) and that from the diffusion equation (\ref{rthree}) for the case of $\eta/s = 0.13$.  The velocity flow obtained from the sound equation has a similar structure as the perturbed energy density with a well defined Mach cone trailing behind the moving parton.  As mentioned before, the velocity flow obtained from the diffusion equation does not describe sound propagation but instead describes the source pulling matter along in its wake.  For our choice of parameters the collective flow induced by the diffusive wake has a longitudinal velocity of approximately $0.15 c$ at a distance of 4 fm behind the source and a much smaller transverse velocity.

In summary, we have derived the pattern of energy and momentum deposition by a fast parton traversing a quark-gluon plasma in perturbative QCD.  Our result depends on the following parameters: the source strength $\alpha_s Q_p^2$, the Debye mass $m_{\rm D}$, the sound velocity $c_s$, and the sound attenuation length, $\Gamma_s$.  Treating the propagation of the disturbance created by the projectile in linearized viscous hydrodynamics we have shown that the fast moving parton excites a sonic Mach cone and a diffusive wake. The intensity of the Mach cone was found to decrease with growing kinematic viscosity.

{\it Acknowledgments:} This work was supported in part by the U.~S.~Department of Energy under grant DE-FG02-05ER4136. JR was supported in part by the NSERC of Canada and BMBF of Germany.

\end{document}